\documentstyle[12pt]{article}
\begin{document}

\title
{\Large \bf Simple  Driven  Maps As  Sensitive  Devices }

\author{ Changsong Zhou$^1$ and  C.-H. Lai$^{1,2}$ \\
        $^1$Department of Computational Science\\
        and $^2$Department of Physics\\
        National University of Singapore,
        Singapore 119260}

\date{}
\maketitle

\begin{center}
\begin{minipage}{14cm}

\centerline{\bf Abstract}
\bigskip

Sensitive dependence of nonlinear systems on initial conditions or
parameters can be useful in applications. We propose in this 
paper that bubbling behavior in simple driven symmetrical  maps 
may be used as a working principle of sensitive 
devices. The system is stable when there is no input and displays 
bursting  behavior  when there is small input. The symmetrical property
of the  bursting  pattern  is very sensitive to the bias of the noisy 
inputs, which makes the system  promising for detecting weak signals 
among noisy environment.

PACS number(s): 05.45.+b;

\end{minipage}
\end{center}

\newpage

A common property of many nonlinear systems is their sensitive
dependence on initial conditions or parameters. This effect can be 
useful in applications. For example, the sensitivity of a chaotic system 
can be used to control its state to  unstable periodic orbits embedded in it 
[1], in targeting the state of the system to  desired points in the state space[2], 
to control the system to follow  a desired  goal dynamics in order to synchronize 
with another system[3] or to  allow a message being encoded in a chaotic series
for the purpose of secure  communication[4], only  by small modifications
of the parameters or state of the chaotic system. The capability of achieving
quite different behavior by applying only small perturbations improves greatly 
the flexibility of a system to be used in various applications. 

By definition, sensitivity is referred to as the growth of small perturbations
to the system. So, naively, sensitivity of nonlinear systems can be used to 
design  sensor devices. Many systems possess a period-doubling bifurcation
when some parameter is varied. Near the onset of a period-doubling bifurcation,
any dynamical system can be used to amplify perturbations near half the fundamental
frequency[5]. One disadvantage associated with the application of such
parameter sensitivity for sensor device purpose is that the control parameter
of the system must be located extremely  close to the critical value of the
bifurcation. 

Recently,  B\"{o}hme and Schwarz proposed to use two identical chaotic systems 
to construct sensitive devices[6]. In particular, they employed the following
symmetrically coupled chaotic systems
\begin{eqnarray}
\dot{x}&=&f(x)-k(x-y)+s_{in},\\
\dot{y}&=&f(y)+k(x-y)-s_{in},
\end{eqnarray}
named chaotic bridge as a sensor device. $s_{in}$ represents a constant
input to be sensed.  The coupling gain $k$ is chosen near the threshold $k_c$ of 
synchronization, so that for $s_{in}=0$, the coupled systems are in
synchronization state, and the output $s_{out}=||x-y||=0$; while for 
$s_{in}\neq 0$, the symmetry of the chaotic bridge is broken, and it may
have a large output at some moment.  Since the synchronization 
manifold is transversely stable for $s_{in}=0$, there must exist
local instabilities in the system in order to  obtain amplification
 of small perturbations. As pointed out by the authors in [6], in the
neighborhood of the boundary $k_c$ of synchronization, one can expect the 
highest sensitivity of the system.

We would like to highlight  the connection of  
the working principle of the above device to 
the phenomenon of {\sl attractor bubbling} studied recently[7-10]. 
When $k$ is just beyond
the threshold $k_c$, the synchronization manifold is 
transversely stable. However, there  still exist  some invariant 
sets, such as the unstable periodic orbits embedded in
the synchronization manifold, which are transversely unstable. 
As a consequence, small perturbations in the systems which destroy its 
synchronization manifold will result in large intermittent bursts from the 
synchronization manifold, no matter how small the perturbations are[8].
This is the origin of the sensitivity of the above system. The difficulties
of  application of the system for sensor devices lie in  practical implementations.
Just like additive  perturbations, any parameter mismatches between 
the systems can also lead to intermittent bursts. 
Parameter mismatches  are inevitable in experiment implementations.
This is the reason that intermittent
desynchronization was observed  beyond the threshold of synchronization
in many experiments of synchronization between well matched electrical
circuits[7-10].  
This effect imposes great difficulties in the experimental implementation of
the above sensor devices, because inevitable parameter
mismatches   lead to large output even for $s_{in}=0$.
The above devices can work only if the
two systems are {\sl ideally identical}, which is extremely  difficult to realize.
On the other hand, small external noise can also result in large bursts when 
$S_{in}=0$, which makes it very difficult to tell a  signal from noise which is
always present in the practical environment.

To avoid the above difficulties, we propose in the following to use  simple driven 
systems  as  sensor devices.

Attractor bubbling and on-off intermittency[11,12] are  common 
behaviors that  occur in coupled nonlinear systems which possess an invariant
manifold. They  can be  achieved in very simple parametrically  driven one-dimensional
maps[12]
\begin{equation}
y_{n+1}=z_nf(y_n),
\end{equation}
where $f(0)=0, \partial f(y)/\partial y|_0\neq 0,$ and $z_n=ax_n>0$ is random or chaotic
driving signal with density function $\rho_z$ and $a$ is a parameter. For
the purpose of the application of the systems as sensitive devices, we require
the maps to have odd symmetry, i.e. $f(-y)=-f(y)$. 

The stability of the invariant manifold $y=0$ is governed 
by the linear equation
\begin{equation}
y_{n+1}=z_ny_n, 
\end{equation}
which describes the evolution of small perturbations transverse to the
invariant line $y=0$.
Here  $\partial f(y)/\partial y|_0$ is absorbed into the parameter $a$. 
The transverse Lyapunov exponent $\lambda$ of the invariant manifold defined
as 
\begin{equation}
\lambda=\lim\limits_{N\to \infty} \frac{1}{N} \sum \limits_{n=1}^{N} \ln z_n
=<\ln z>
\end{equation}
determines the stability of the invariant manifold. The critical point 
$a_c$ at which $\lambda=0$ is the onset point of on-off intermittency[12]. 
For $z_n$ being a uniform random driving signal on  $(0, a)$, $\rho_z=1/a$, and 
 $<\ln z>=\ln a -1=0$ gives $a_c=e$. 
Just above the onset point, $a\geq a_c$, the random driven system displays universal 
features of  on-off intermittency behavior, which are unaffected by the
form of the confining nonlinearity[12]. 
The nonlinearity of the system serves to bound or reject the 
dynamics back towards small values of $y$ after bursts. 
For $a$ below $a_c$, the invariant manifold $y=0$ is stable, but the stability is 
quite weak if $a$ is near $a_c$. Attractor bubbling     
occurs in the system when there are inputs  of perturbations  such as noise.

For a sensitive device, it should be stable when there is no input, and  is
expected to produce large outputs when there are small perturbations.
In this paper, we are not considering the on-off intermittency behavior 
with $a\geq a_c$. For the  purpose of sensor purpose, we employ the
bubbling behavior with $a<a_c$.  
The sensor system  reads
\begin{equation}
y_{n+1}=z_nf(y_n)+s_{in},
\end{equation}
and $s_{out}=y$. When there is no input, i.e.  $s_{in}=0$, $y=0$ is a stable solution,
and the output $s_{out}=0$. 
Since the critical parameter $a_c$ and the evolution of small perturbations are
independent of the form of the  nonlinearity,  one can choose a map which is simple
for implementation. For example, we employ a piecewise linear map
\begin{equation}
f(y)=\left\{
\begin{array}{ll}
\frac{c_1}{c_2}(-c_1-c_2-y),& y<-c_1,\\
y,                         & |y|\leq c_1,\\
\frac{c_1}{c_2}(c_1+c_2-y),& y>c_1,\\
\end{array}\right.
\end{equation}
where $c_1$ and $c_2$ are two parameters, 
as shown in Fig. 1.

If there is no input, $s_{in}=0$, staring from a small initial condition $y_0=p$,
we have $y_n=z_{n-1}z_{n-2}\cdots z_0p=k_np$. The average order of the factor $k_n$,
which can be  defined as $<\ln k_n>$, decreases linearly with $n$ as
\begin{equation}
<\ln k_n>=n<\ln z>= n(\ln a -1). 
\end{equation}
Now, suppose there is  a small constant input $s_{in}=p$, starting from $y_0=0$,  the evolution 
of the output reads
\begin{equation}
\begin{array}{c}
y_1=(z_0+1)p=k_1p,\\
y_2=(z_1z_0+z_1+1)p=k_2p,\\
\cdots\\
y_n=(z_{n-1}\cdots z_0+\cdots+z_{n-2}z_{n-1}+z_{n-1}+1)p=k_np, 
\end{array}
\end{equation}
if $\max (|y_n|)=|p|(1+a+\cdots a^n)=|p|\frac{a^{n+1}-1}{a-1}\leq c_1$, or,
$n\leq n_c={\hbox {int}}[\ln (c_1(a-1)/|p|+1)/\ln a]-1$
, where ${\hbox {int}}[x]$ is the interger part
of a real number $x$. The average order of $k_n$, $<\ln k_n>$, cannot be 
obtained analytically as that for $s_{in}=0$ in Eq. 8. A numerical estimation 
 of $<\ln k_n>$  is carried
out with $10^6$ samples of $k_n$.  
Unlike the case $s_{in}=0$, it is an increasing function of $n$, 
as shown in Fig. 2. 

When $n>n_c$, $y_n$ has nonvanishing probability
to exceed $y_n>c_1$. Following B\"{o}hme and Schwarz, a measure $s$ for the sensitivity of the system to constant input can be defined as
\begin{equation}
s=\frac{\max\limits_{n}|y_n|}{|s_{in}|}.
\end{equation} 
Since $\max(k_n)=\frac{a^{n+1}-1}{a-1}$ increases exponentially with $n$, an infinitely small
input value can create a finite output. 
 The largest output  $\max\limits_{n}|y_n|=ac_1$ is due to the confinement of the nonlinearity of the map.
The sensitivity $s$ of the system thus goes to infinity and is referred to as {\sl
supersensitivity}[6].

An important question concerning the system is  the time needed for small input
to produce a large output.  We examine the time $N$ for
a small input $p$ to produce for the first time  an output $s_{out}\geq c_1$.  
The distribution $P(N)$ of $N$ is
the following  probability
\begin{equation}
P(N)={\hbox {Prob}}(\bigcap\limits_{i=1}^{N-1}y_i<c_1\bigcap y_N\geq c_1).
\end{equation}
By defining the event
$E_N=\bigcap\limits_{i=1}^{N}y_i<c_1$ and the corresponding probability $\Lambda_N={\hbox {Prob}}(E_N)$, it follows that
$$P(N)=\Lambda_N-\Lambda_{N-1},$$
which is a function of both $a$ and $p$.
In principle, $\Lambda_N$ can be evaluated by the joint density  $\Phi(K)$ of $k_1, k_2, \cdots, k_N$,
namely, 
$$\Phi(K)=\frac{\rho(z_0)\rho(z_1)\cdots \rho(z_{N-1})}{|J|}=\frac{\rho(k_1-1)\rho(\frac{k_2-1}{k_1})\cdots \rho(\frac{k_{N}-1}{k_{N-1}})} {k_1k_2\cdots k_{N-1}},$$
where $J$ is  the Jacobian of the transformation between $k_i$'s and $z_i$'s defined in Eq. (9). 
Specifically, one has
$$
\Lambda_N=\int _1^b dk_N \int_{(k_N-1)/a}^{b} \rho(\frac{k_{N}-1}{k_{N-1}}) \frac{dk_{N-1}}{k_{N-1}}\cdots
\int_{(k_{i+1}-1)/a}^{b} \rho(\frac{k_{i+1}-1}{k_i}) \frac{dk_i}{k_i}\cdots \nonumber $$
$$\times \int_{(k_{2}-1)/a}^{b} 
\rho(\frac{k_{2}-1}{k_1}) \rho(k_1-1) \frac{dk_1}{k_1}$$
where $b=c_1 /p$ if $i>n_c$ and $b=\frac{a^{i+1}-1}{a-1}$ if $i\leq n_c$. 
However, to the best of our knowledge, a  closed-form solution of  the above integral for any $N$ is not 
available.

In the following, we are going to carry out some simulations.  We specify $c_1=1$ 
and $c_2=2$ in these simulations.
In Fig. 3, as an example, the output sequence is shown for $s_{in}=1.0\times 10^{-4}$ at $a=2.6$. 
The dashed lines indicate the switch on and off of the constant input. 
The output in the presence of input is
a intermittent process, similar to the result of the chaotic bridge in [6].  

In the  next simulation, we estimate $P(N)$ for different values of $a$ and $p$, as shown in Fig. 4. 
It is seen that the distributions peak at 
rather small $N$ values, and after the peak, they  decrease exponentially. The
average time $<N>$ for first putting out  $y_N \geq c_1$  is also evaluated as a function 
of $a$ and $p$ 
in Fig. 5(a) and (b), respectively.   
So,  on average, the closer the $a$ to $a_c$ and the larger the input  $p$, the
quicker the system reaches a large output. 

The simulation results show that this system may be used as a detector for
weak signal. However, the above discussion is only valid in a noise-free
 environment.   In the practical application of the system as a detector,  external noise 
is unavoidable. The system now reads  
\begin{equation}
y_{n+1}=z_nf(y_n)+s_{in}+e_n,
\end{equation}
where $e_n$ denotes external noise. It is plausible to assume that 
$e_n$ has vanishing mean value and a Gaussian distribution $\sigma N(0,1)$,
with a standard deviation $\sigma$. The behavior of the system  in the 
presence of noise is quite
different from the noise-free case, because bubbling occurs even without a signal $s_{in}$. 
Very small external noise can also lead to large output of the system, as 
illustrated in Fig. 6 (a) with $a=2.6$ and $\sigma=1\times 10^{-4}$.
In this system with  $a<a_c$, 
bursting behavior always means that there are some inputs to
the system, and the largest output will be independent of the inputs.
The problem  now becomes whether we can  distinguish  that the input is a
meanful signal or just  noise, and moreover  whether we can  detect any significant 
signal among the white noise environment. As will be  shown in the
following, this system is
quite promising for this task, because the symmetrical property of the
bursting behavior is very sensitive to the bias of the  inputs.

This sensitivity is due to the odd symmetry of the map.
An inspection of Fig. 6(a) reveals that the number of large bursts to 
positive and negative values is quite symmetric for inputs of white 
noise. When a small positive constant input $s_{in}=p=0.3\times 10^{-4}$ 
is present along with the noise, the symmetry is clearly broken, as seen
in Fig. 6(b). Note that the constant input $p$ is much smaller than the noise level
$\sigma$ in this example. This result indicates that the symmetry of burst is quite
sensitive to the bias of the total inputs $s_{in}+e_n$ of the system.
To characterize the symmetry breaking property quantitatively, we introduce 
the degree of asymmetry $D_{asy}$ as
\begin{equation}
D_{asy}=\frac{N_+-N_-}{N_++N_-},
\end{equation}
where $N_+$ ($N_-$) is the number of large burst ($|y|\geq 1$) to positive
(negative) values during  a period of observation time $T$. Because of  the symmetry of the
map and that of the white noise, one can expect that
$D_{asy}\approx 0$ for white noise inputs. For positive (negative) constant inputs
in the noise-free case, it is clear that $D_{asy}=1 (-1)$. 

A comparison between Fig. 6(a) and (b) also shows that large bursts occur more frequently when
constant input is present with the noise. We define the bursting frequency $F$ as
\begin{equation}
F=\frac{N_++N_-}{T}.   
\end{equation}
We expect that $F$ increases with larger constant input $p$ 
and larger noise level $\sigma$.

A  measure of the significance of a constant signal among the
noise can be the ``signal to noise ratio'' $R=p/\sigma$. In the following, simulations are
carried out to examine the dependence of $D_{asy}$ and $F$ on $R$ for
different noise level $\sigma$  and system parameter  $a$. The results shown in Fig. 7 are obtained
with $T=2\times 10^6$. 
The result of $D_{asy}$ is very interesting: as a function of $R$,
$D_{asy}$ is independent of  the noise level $\sigma$ and is not sensitive to parameter $a$. Whether a
constant signal embedded in the noise environment can be detected  depends only
on its significance with respect to the noise level.  
However, $a$ and $\sigma$ have
effect on the bursting frequency $F$, as seen from Fig. 7(b).       
A combination of Fig. 7(a) and (b) makes it possible to determine the
noise level and $R$, and thus to estimate the amplitude of the constant signal in a
observation.  

To get a good estimation of $D_{asy}$ and $F$, $T$ should be large enough. In
practice, one may not expect a signal that stays constant for so long a time.
The term {\sl constant signal} is a concept  relative to the time scale of the
detector, and the time scale can be controlled in the 
implementation. In numerical experiment of this dimensionless system,  
we can simulate a shorter signal (or a  ``slower'' detector)
with smaller  $T$, e.g., $T=2000$. In this case, $D_{asy}$ has large
fluctuations, as shown in Fig. 8(a) where $D_{asy}$ of 20 realizations of the 
driving $z_n$ and white noise $e_n$  of the system  are plotted for each $R$ value. 
When $R$ is getting larger, more points coincide at $D_{asy}=1$.
A good way to examine the fluctuation behavior is to
construct a histogram of $D_{asy}$, as shown in Fig. 8(b) for $a=2.6,
\sigma=10^{-4}$. The results show  that even for quite short signal
and low $R$ value, $D_{asy}$ has very high probability near $D_s=1$.
An implication of the results is that
several detectors can be used at the same time to detect  and confirm a short
and weak signal embedded in white noise.   

As an example of a  little more realistic input signal, we present the
response of the system to the noisy input $A\sin (0.003n)+e_n$, where
the noise level is $\sigma=1\times 10^{-4}$ and the amplitude of the sine
wave is $A=0.3\times 10^{-4}$. Both the total input and the output of the
system are displayed in Fig. 9. The bursting  feature reflects the weak wave 
among the noise quite clearly.

In summary, we have shown that a kind of very simple driven symmetrical maps below the 
onset point of on-off intermittency have two distinguishing features of  (i)
being  stable at the invariant state $y=0$ and (ii) being sensitive to small input.  
In practice, the environment cannot be noise-free, and the systems exhibit 
bubbling behavior in the presence of noise. Another interesting  and useful property 
of the systems is that the bursting pattern is symmetrical for white noise input, and 
the  symmetry  is broken when there is signal among the noise environment. The significance
of the signal is manifested by the degree of asymmetry in the bursting pattern.   
These features  make them  promising candidates for designing  sensitive
devices.

Although our study is based on numerical simulations of a mapping model, it should be 
noted that system response to small inputs is governed by its linearized equation, and the nonlinearity 
only serves to keep the system bounded.  
Many long time properties  shown above thus are universal  in a  class of driven  systems  possessing odd 
symmetry. The following can be advantages for such systems when considering applications in sensitive devices: 

1) The sensitivity is maintained in a large range of parameter below the critical
point. This  avoids the difficulty of locating parameter in a very small
neighborhood of a bifurcation point in a period-doubling system.

2) The sensitivity of the system is, in principle,  infinite. In a noise-free environment,
  an infinitesimal input signal  can produce a finite
output.  When noise is present, weak  signal can also be manifested by the asymmetry in the bursting
pattern. The degree of asymmetry depends on the significance of the signal with respect to the noise. 
The sensitive behavior is universal for different
forms of nonlinearity of the systems, as well as for different form of driving signals.
This is very useful because one can thus choose a system that is  
simple and easy to implement in practice.

It could be meaningful to consider implementation of such simple systems and explore their
application in small signal detection. Since in applications, pivotal role is 
played by the symmetry  properties of the system, one should take care to maintain such properties. 
In order to avoid perturbations  which may make the system appreciably asymmetric, one 
should avoid using different parameters
for the two symmetrical parts of the systems.   Also, one should note that it  takes longer for the 
system to produce static output states for lower level of inputs.  There seems to be a frequency cutoff 
associated with the input levels and the relaxation time of the systems. Above the cutoff, the small 
signal in the noise can no longer be manifested by  clear asymmetry in the bursting pattern. Such 
limits should be taken into consideration in applications.

\bigskip
{\bf Acknowledgements:}

This work was supported in part by research grant RP960689 at the National
University of Singapore.  CZ was supported by  NSTB.

\newpage

{\bf References}
\begin{description}
\item [1.] E. Ott, C. Grebogi, and J. Yorke, Phys. Rev. Lett. {\bf 64}, 1196(1990). 
\item [2.] T. Shinbrot, E. Ott, C. Grebogi, and J. Yorke, Phys. Rev. Lett. {\bf 65}, 3215(1990);
 T. Shinbrot, {\sl et al}, Phys. Rev. Lett. {\bf 68}, 2863 (1992).
\item [3.] Y. C. Lai, C. Grebogi, Phys. Rev. E {\bf 47}, 2357 (1993). 
\item [4.]S. Hayes, C. Grebogi, and E. Ott, Phys. Rev. Lett. {\bf 70}, 3031(1993);
S. Hayes, C. Grebogi, E. Ott, and A. Mark,  Phys. Rev. Lett. {\bf 73}, 1781(1994);
E. M. Bollt and M. Dolnik, Phys. Rev. E {\bf 55}, 6404 (1997).
\item [5.] K. Wiesenfeld and B. McNamara, Phys. Rev. Lett. {\bf 55}, 13(1985).
\item [6.] F. B\"{o}hme and W. Schwarz, in {\sl Nonlinear Dynamics of Electronic Systems},
eds by A.C. Davies and W. Schwarz (World Scientific, Singapore, 1994), pp 281. 
\item [7.] P. Ashwin, J. Buescu, and I. Stewart, Phys. Lett. A {\bf 193}, 126(1994).
\item [8.] J. F. Heagy, T. L. Carroll, and L. M. Pecora,
 Phys. Rev. E {\bf 52}, R1253(1995).
\item [9.] D. J. Gauthier and J. C. Bienfang, Phys. Rev. Lett. {\bf 77}, 1751(1996).
\item [10.]S. C. Venkataramani, B. R. Hunt, E. Ott, D. J. Gauthier, and J. C. Bienfang,
            Phys. Rev. Lett. {\bf 77}, 5361(1996). 
\item [11.] N. Platt, E. A. Spiegel, and C. Tresser, Phys. Rev. Lett. {\bf 70}, 279
(1993).
\item [12.] J. F. Heagy, N. Platt, and S. M. Hammel, Phys. Rev. E {\bf 49}, 1140
(1994).
\end{description}

\newpage
{\large \bf Figure Captions}
\begin{description}

\item Fig. 1 The piecewise linear map $f(y)$. Here $c_1=1$ and $c_2=2$.
\item Fig. 2 $<\ln k_n>$, the  average order of $k_n$,  as a function of $n$.  
             It decreases linearly for the case without input (plot a) and increases
             for the case with input (plot b).
\item Fig. 3 An illustration  of the output process of the sensitive system at $
              a=2.6$. The constant input is $p=10^{-4}$, and is switched on
             and off alternately for every 500 iterations, as shown by the dashed lines.  
\item Fig. 4 Numerically evaluated distribution  of $N$.  The parameters of the plots are:
            (a) $a=2.5, p=10^{-3}$, (b) $a=2.5, p=10^{-4}$ and (c) $a=2.6, p=10^{-4}$. 
\item Fig. 5 (a) Average value of $N$ as a function of $a$ with  $p=10^{-3}$, and (b) 
             as a function of $p$  with $a=2.6$. 
\item Fig. 6 (a) Output of the system when only white noise is present as input. 
             (b) Output of the system when constant signal $p=0.3\sigma$ is
                 present along with the noise. The parameters are $a=2.6, \sigma=1\times 10^{-4}$. 
                 Noting the change of symmetrical property of the bursting pattern of the system.
\item Fig. 7 (a) Degree of asymmetry $D_{asy}$ as a function of ``signal to noise ratio'' $R$ for
different $a$ and $\sigma$.
             (b) Bursting frequency $F$ as a function of $R$ for different $a$ and $\sigma$.
               (b) shares the same legends of (a).
              The results are obtained from observation during  a period of time $T=2\times 10^6$.
              
\item Fig. 8 (a) $D_{asy}$  of short time observation, $T=2000$. (b) Normalized histograms of
$D_{asy}$ for different $R$ value. The histograms are constructed with 50000 observations for each
$R$ value. The parameters are $a=2.6, \sigma=1\times 10^{-4}$.

\item Fig. 9 (a) A weak sine wave embedded in the noise. (b) The response of the detector to the
inputs of (a).  

\end{description}

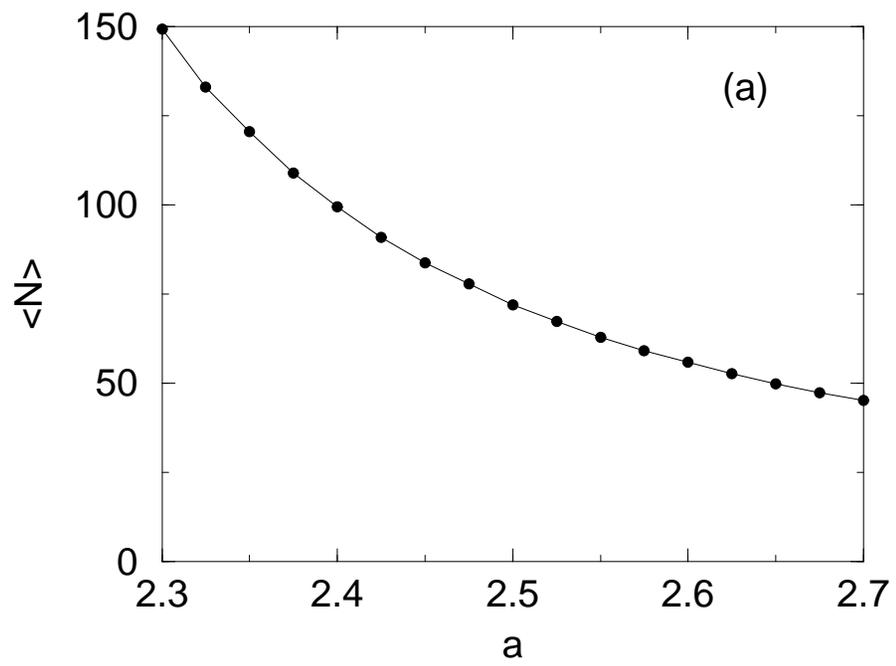
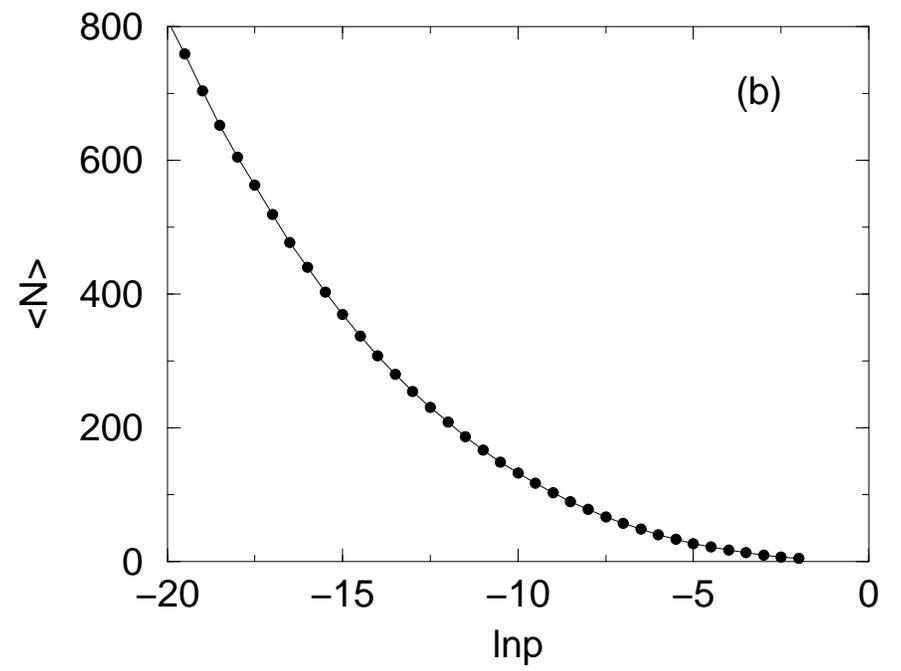
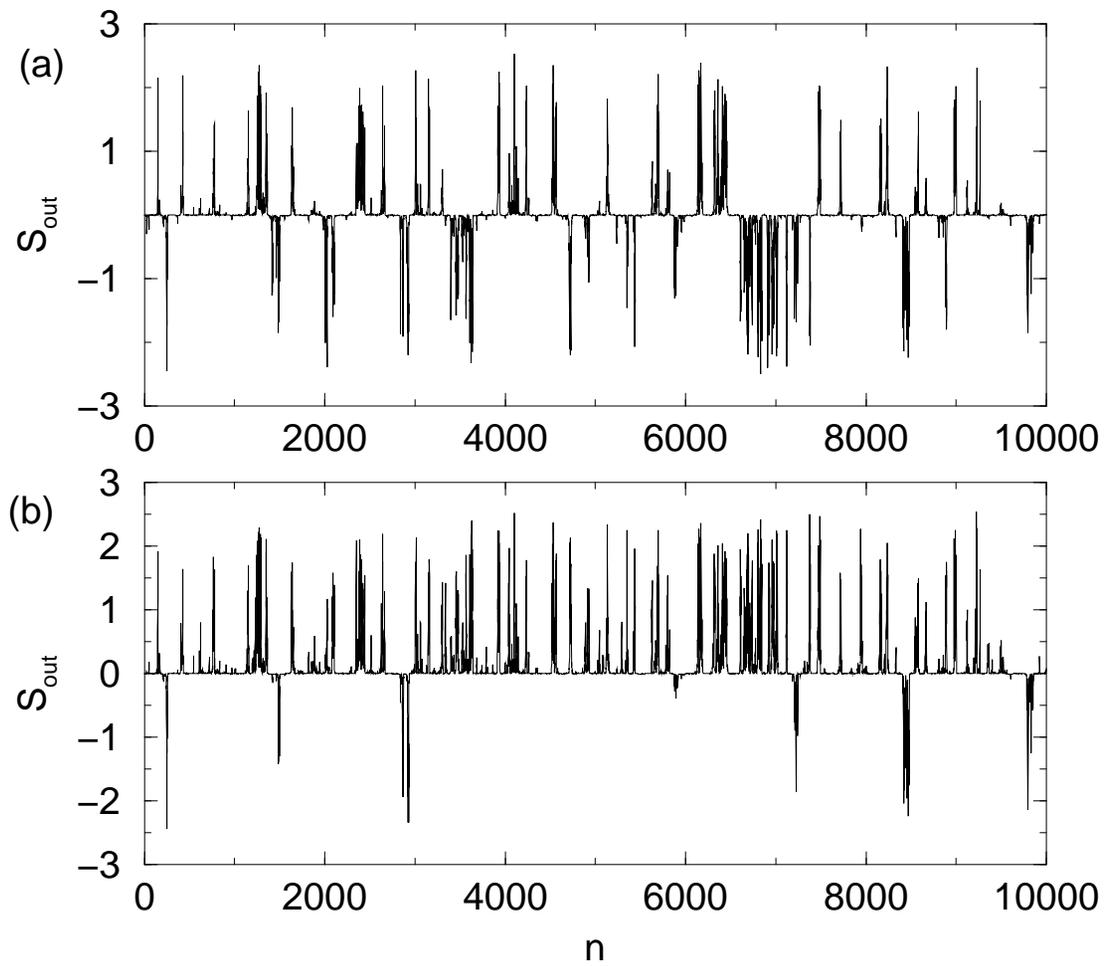
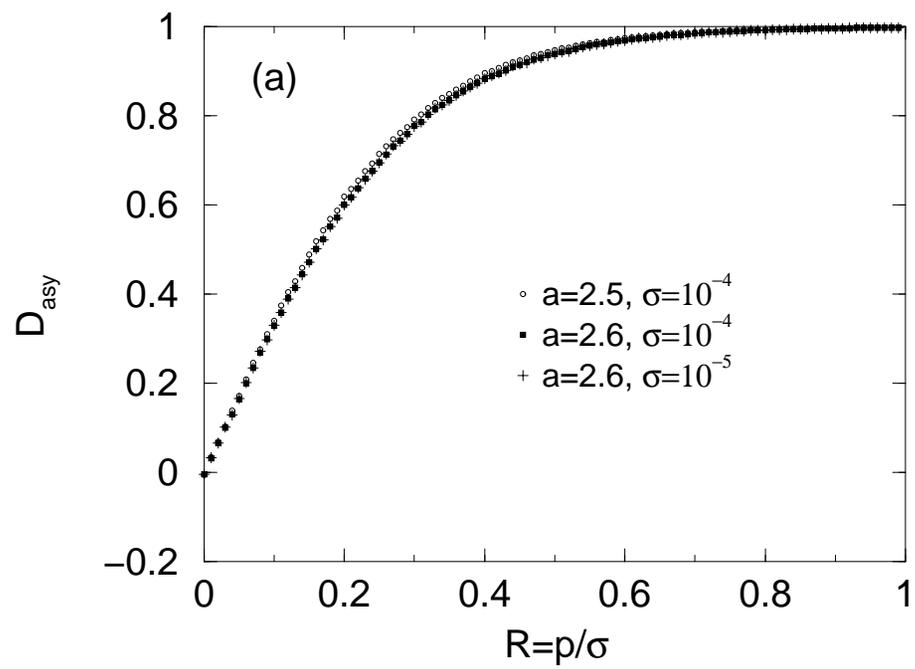
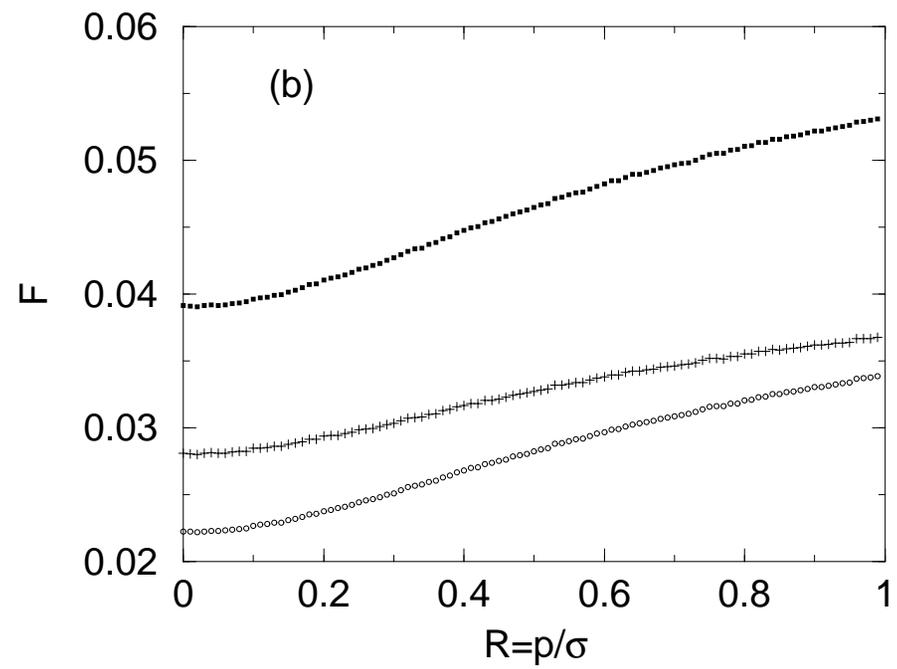
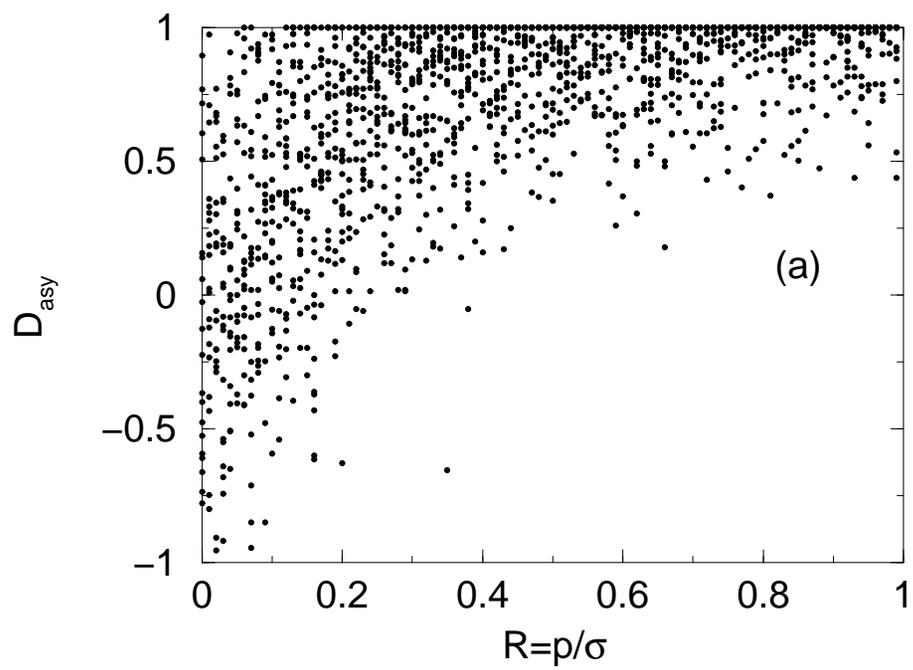
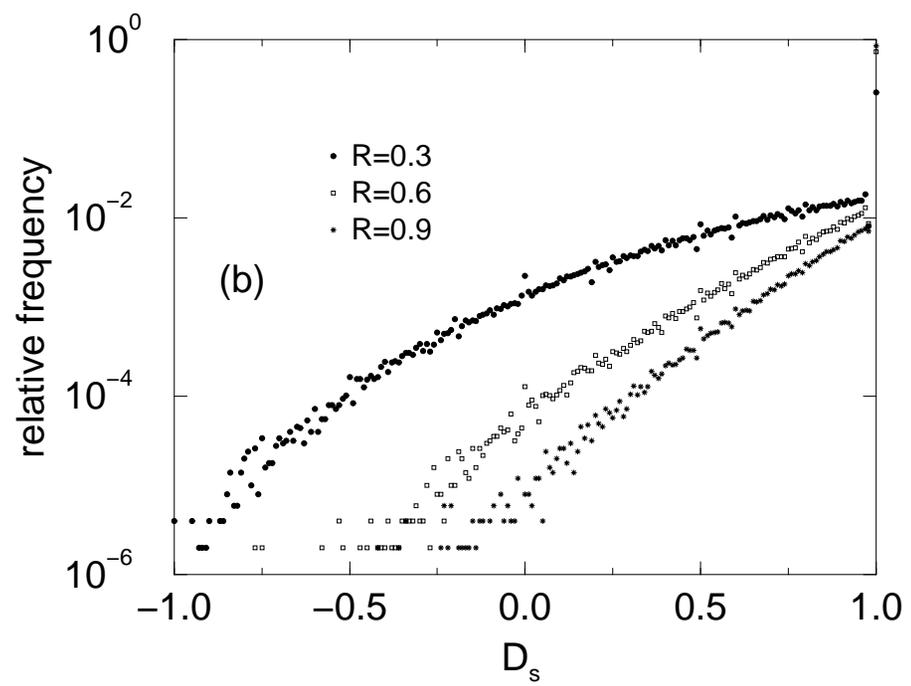
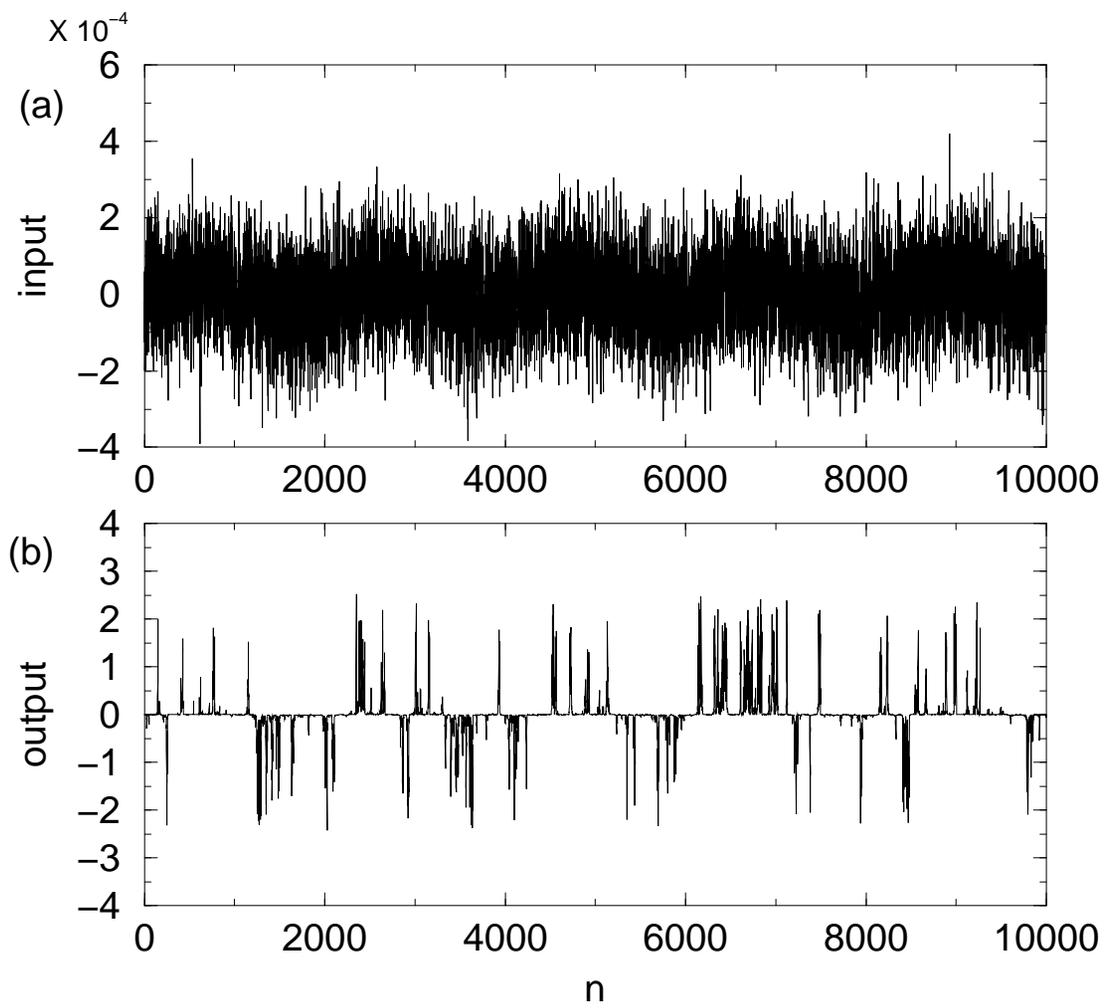
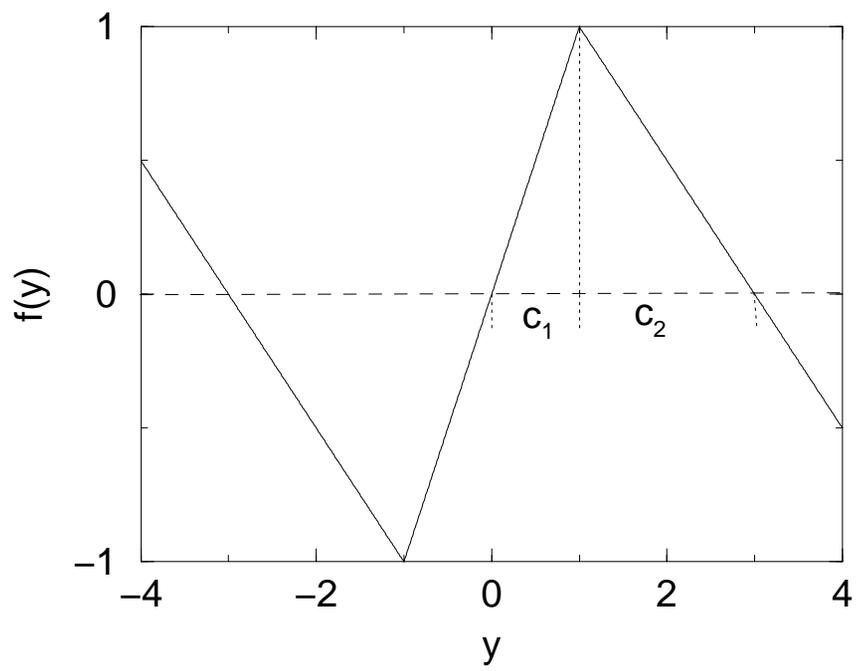
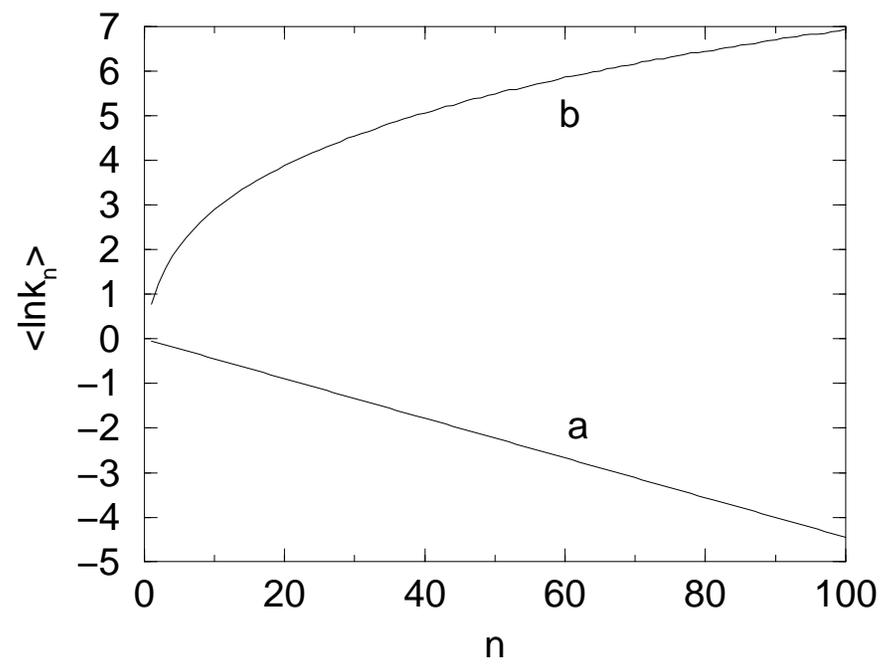
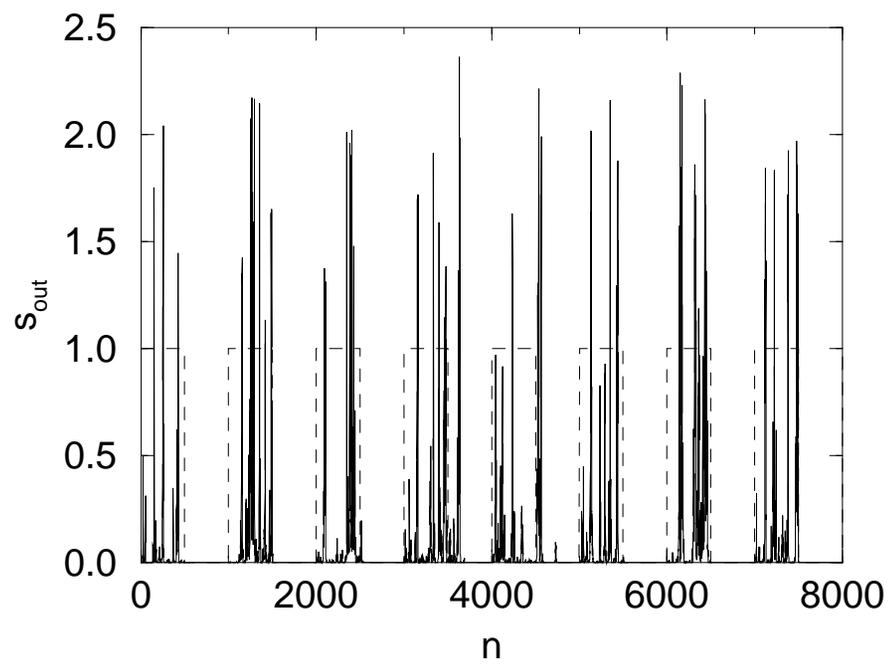
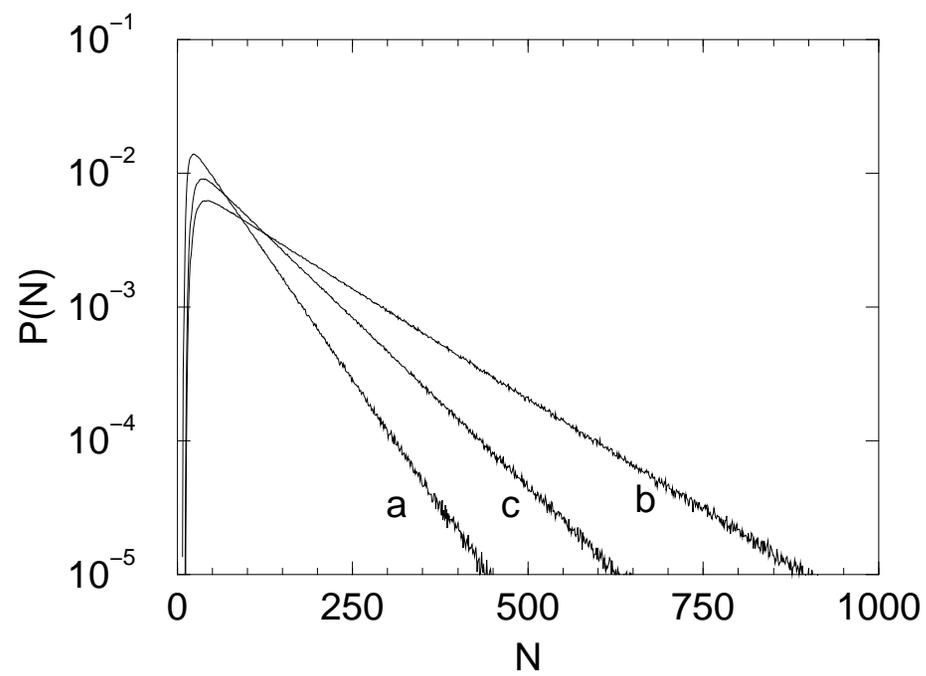
\end{document}